\documentclass[sigconf, nonacm, natbib=false]{acmart}
\AtBeginDocument{%
  }

\setcopyright{none}
\copyrightyear{2026}
\acmYear{2026}
\acmDOI{}
\acmConference[ACM Creativity \& Cognition XAIxArts]{Explainable AI for the Arts Workshop 2026}{July 13, 2026}{London, UK}
\acmISBN{}

\RequirePackage[
  datamodel=acmdatamodel,
  style=acmnumeric,
  sortcites=true,
  backend=biber,
  maxbibnames=3,   
  minbibnames=2,
  doi=true,       
  isbn=false,      
  url=false        
]{biblatex}
\AtEveryBibitem{
  \clearfield{urlyear}
  \clearfield{note}
  \clearfield{addendum}
  \iffieldundef{doi}{}{\clearfield{url}} 
}

\renewcommand\footnotetextcopyrightpermission[1]{} 

\begin{document}

\title[Real-Time AttentionBender]{Real-Time AttentionBender: Granular Interactive Network Bending of Video Diffusion Transformers}


\author{Adam Cole}
\email{a.cole@arts.ac.uk}
\orcid{0000-0001-9715-314X}
\affiliation{%
  \department{Creative Computing Institute}
  \institution{University of the Arts London}
  \city{London}
  \country{UK}
}

\author{Rebecca Fiebrink}
\email{r.fiebrink@arts.ac.uk}
\orcid{0000-0002-7609-2234}
\affiliation{%
  \department{Creative Computing Institute}
  \institution{University of the Arts London}
  \city{London}
  \country{UK}
}

\author{Mick Grierson}
\email{m.grierson@arts.ac.uk}
\orcid{0000-0002-6981-5414}
\affiliation{%
  \department{Creative Computing Institute}
  \institution{University of the Arts London}
  \city{London}
  \country{UK}
}


\begin{abstract}
Generative video models have achieved remarkable visual fidelity, yet their prompt-only interface offers thin creative agency and obscures the model's material process from the artists working with it. We present \textit{Real-Time AttentionBender}, a tool that extends the practice of network bending across the full depth of the video diffusion transformer (DiT) and brings it into live, interactive generation. Built as a plugin within the DayDream Scope ecosystem and wrapping open-source real-time Wan pipelines, the tool exposes self-attention, cross-attention, and the feed-forward network as independently manipulable surfaces, with targeting down to individual diffusion steps, DiT layers, prompt tokens, and hidden neurons. The immediacy of live manipulation affords what we call \textit{material intimacy} with the model: a responsive, near-mechanistic feel for how specific layers and neurons shape generated video. We position the tool as simultaneously an explainable AI probe into transformer internals and an expressive instrument for discovering aesthetics outside the model's default representational space.
\end{abstract}




\begin{teaserfigure}
  \includegraphics[width=\textwidth]{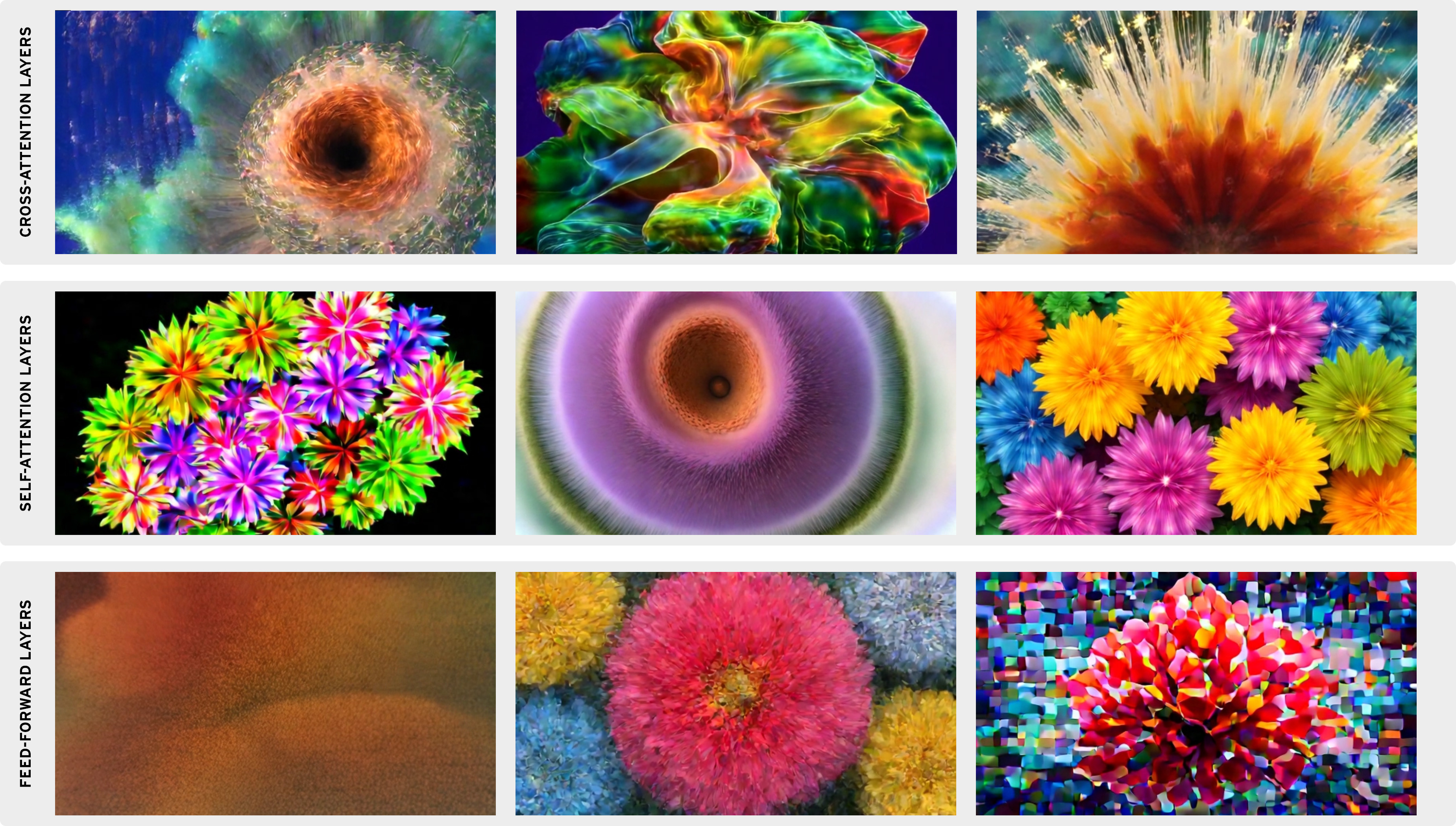}
  \caption{\textit{Real-Time AttentionBender} Expressive Potential: This diverse set of samples was generated with the exact same video model and text prompt, but with varying attention bending settings. Each row shows modulations targeting a single DiT layer type (top to bottom: cross-attention, self-attention, feed-forward).}
  \Description{Enjoying the baseball game from the third-base
  seats. Ichiro Suzuki preparing to bat.}
  \label{fig:teaser}
\end{teaserfigure}


\maketitle

\section{Introduction}

Generative video models have achieved striking quality and temporal consistency, pushing the boundaries of synthetic media production. However, to the working artist and creative coder, these massive architectures remain deeply opaque. While generative outputs are increasingly realistic, the ubiquitous prompt-only interface severely limits creative agency. Text prompts restrict artists' ability to build intuition for the model's material process, or to meaningfully intervene and work beyond the network's default representational tendencies. If artists are to co-create \textit{with} these systems rather than simply feed them prompts, it is essential we expose the black box of AI video models through direct, responsive interaction.

To address this, we build upon the lineage of Network Bending~\cite{broadNetworkBendingExpressive2021, abuzuraiqExplainabilityinActionEnablingExpressive2025}: the practice of directly manipulating the internal activations of generative models to discover novel visual languages and glitch aesthetics. We previously introduced \textit{AttentionBender}~\cite{adamcoleAttentionBenderManipulatingCrossAttention2026}, a tool that allowed artists to apply 2D transforms (rotation, scaling, translation) to the cross-attention maps of video diffusion transformers (DiTs). While effective for structural manipulation, its scope was limited to offline rendering and isolated to cross-attention layers, breaking the tight feedback loop necessary for instinctual creative exploration.

In this paper, we present \textit{Real-Time AttentionBender}, significantly expanding the scope, depth, and interactivity of the original tool. Built as a plugin within the DayDream Scope ecosystem \cite{daydreamScope2026c} and wrapping open-source real-time Wan pipelines \cite{wanteamWanOpenAdvanced2025}, it delivers live network manipulation at interactive frame rates on a single GPU. To our knowledge, this work represents the first application of network bending to the full depth of the video transformer architecture, and the first time such manipulations can be explored in real-time. Specifically, this tool contributes:
\begin{enumerate}
    \item \textbf{Real-Time Interaction:} A live, responsive interface for modulating generative video diffusion pipelines.
    \item \textbf{Full DiT Block Modulation:} An expansion of network bending controls across all core components of the DiT block: self-attention, cross-attention, and feed-forward layers.
    \item \textbf{Granular Targeting:} The ability to isolate modulations by diffusion step, specific layer, prompt token, or down to the individual hidden neuron.
\end{enumerate} 
By enabling this level of fine-grained, responsive control, \textit{Real-Time AttentionBender} functions simultaneously as an XAIxArts~\cite{bryan-kinnsXAIxArtsManifestoExplainable2025} probe into transformer attention mechanics, and as a highly expressive creative instrument for discovering novel aesthetics outside the model's learned representational space. The immediacy of live manipulation produces what we term \textit{material intimacy}~\cite{adamcoleAttentionBenderManipulatingCrossAttention2026} with the model---a responsive, near-mechanistic feel for how specific layers and neurons contribute to the final generated video.

\begin{figure}[h!]
\centering
\includegraphics[width=1\linewidth]{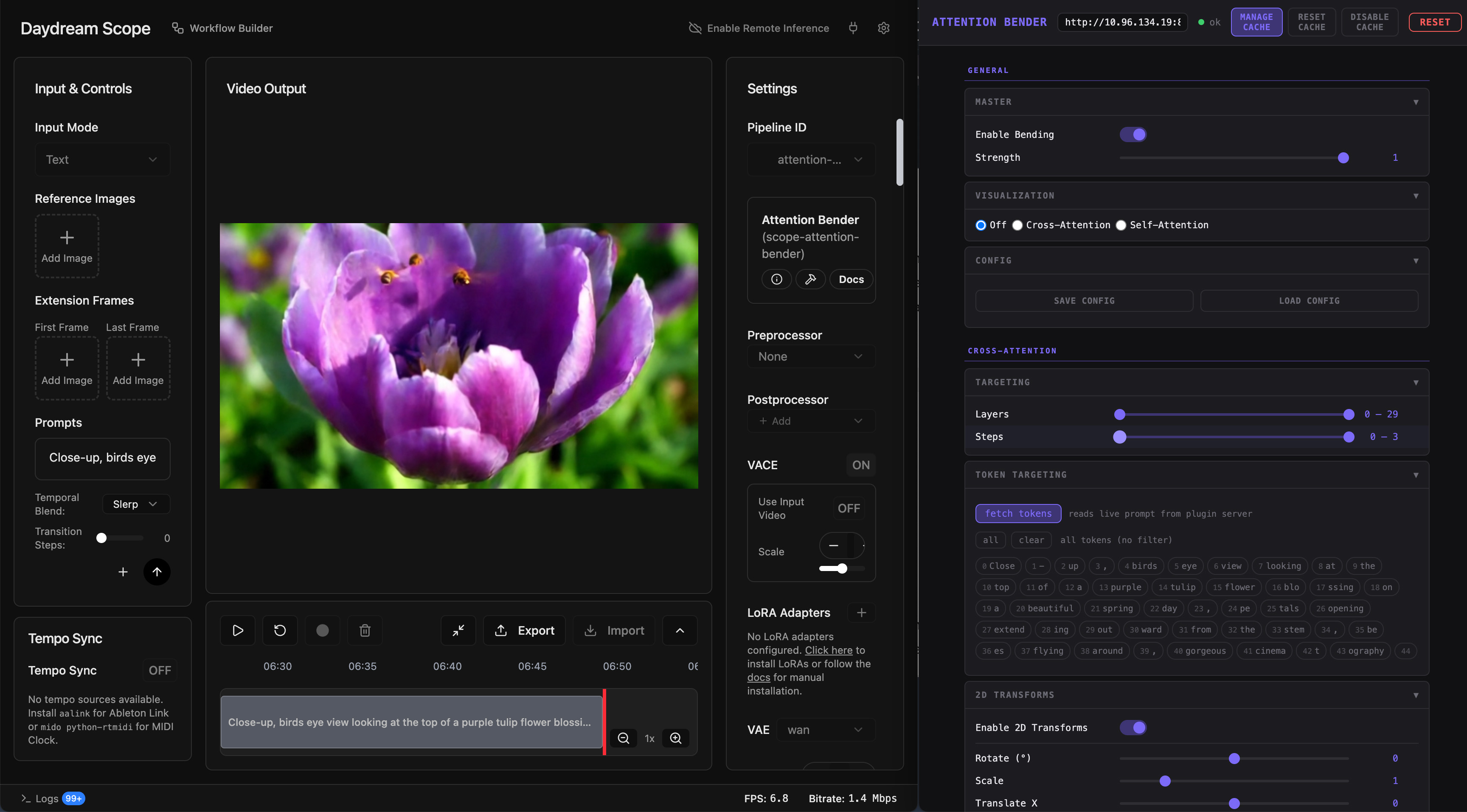}
\caption{Real-Time \textit{AttentionBender} Interface: The left panel displays the live generative video output, while the right panel houses interactive attention bending controls organized by self-attention, cross-attention, and the feed-forward layers.}
\label{fig:interface}
\end{figure}

\section{Background}

\subsection{Explainable AI Video for the Arts}

While state-of-the-art AI video generation models (e.g., Veo \cite{humeMeetFlowAIpowered2025}, Sora \cite{openaiSoraBlogpost2024}, Wan~\cite{wanteamWanOpenAdvanced2025}, and Hunyuan \cite{kongHunyuanVideoSystematicFramework2025}) continue to rapidly advance in both performance metrics and sample fidelity, they severely lack meaningful creative control. The dominant paradigm of prompt-based generation offers limited creative options, completely abstracting away the computational process. This opacity extends to the model itself: as artists, we do not yet possess a material understanding of how this vast collection of learned weights coalesce to produce such a wealth of representations.

To address this, we adopt the methodology of Explainable AI for the Arts (XAIxArts)~\cite{bryan-kinnsXAIxArtsManifestoExplainable2025}. Rather than treating the neural network as a black box that maps text to pixels, XAIxArts advocates for interrogating the internal mechanics of the system. This project embodies this approach by presenting an expressive and controllable interface that exists beyond the prompt box. By allowing artists to probe the system down to the level of individual neurons, we facilitate the development of a richer material intuition for AI video.

\subsection{Video DiTs and Real-Time Architectures}

Recent advancements in generative content are driven primarily by video diffusion models~\cite{hoVideoDiffusionModels2022, blattmannAlignYourLatents2023}, and increasingly, the Video Diffusion Transformer (DiT) architecture~\cite{peeblesScalableDiffusionModels2023, openaiSoraBlogpost2024}. For this project, we build upon Wan~\cite{wanteamWanOpenAdvanced2025}, an open-source latent \cite{rombachHighResolutionImageSynthesis2022} video DiT that has become the foundation for a diverse ecosystem of research. The Wan architecture is elegantly structured as a sequence of DiT blocks, each containing three primary components: a self-attention layer, a cross-attention layer, and a feed-forward network (FFN) with a hidden layer---30 blocks in the 1.3B model, 40 in the 14B.

Historically, diffusion models have been inherently limited in real-time settings because their bidirectional attention mechanism requires joint processing of the full spatiotemporal sequence (including future frames), compounded by a long iterative denoising process. However, recent algorithmic breakthroughs---such as causal attention re-orientation~\cite{yinSlowBidirectionalFast2025}, few-step diffusion distillation~\cite{yinOnestepDiffusionDistribution2024}, and self-forcing~\cite{huangSelfForcingBridging2025a}---have dramatically accelerated these systems. Consequently, Wan now serves as a common backbone for numerous real-time and near real-time AI video pipelines, including CausVid~\cite{yinSlowBidirectionalFast2025}, Self-Forcing~\cite{huangSelfForcingBridging2025a}, LongLive~\cite{yangLongLiveRealtimeInteractive2025}, and Krea~\cite{KreaRealtime14B2025}. This shared backbone gives us a consistent target architecture across the real-time video ecosystem.

Our work addresses a critical gap in the existing creative AI literature. While network bending has been fruitfully applied to older architectures like GANs~\cite{broadNetworkBendingExpressive2021} and U-Nets~\cite{abuzuraiqExplainabilityinActionEnablingExpressive2025}, and while the original \textit{AttentionBender} explored offline manipulations of DiT cross-attention~\cite{adamcoleAttentionBenderManipulatingCrossAttention2026}, there currently exists no framework for exploring the \textit{entire} DiT block interactively. By wrapping these newly performant open-source real-time pipelines, our tool makes live interaction possible across every layer of the transformer, building the instinctual understanding required for an embodied XAIxArts practice.

\section{Real-Time AttentionBender}

\subsection{System Architecture}

\textit{Real-Time AttentionBender} is built as a custom plugin within the DayDream Scope ecosystem. The plugin operates via a dynamic wrapper around existing real-time diffusion pipelines (currently supporting LongLive and Krea). During initialization, the plugin monkeypatches key classes within the Wan DiT blocks---specifically \texttt{WanSelfAttention}, \texttt{WanT2VCrossAttention}, and the \texttt{Feed-Forward Sequential module}. These patched classes are injected with global listeners that fetch user-defined modulation parameters at inference time. This architecture allows us to alter network behavior on the fly without interrupting the generative loop. In our current configuration, wrapping the LongLive pipeline at a resolution of $320 \times 576$, the system achieves a responsive \textasciitilde15 frames per second on a single NVIDIA RTX A6000 Pro GPU. To further aid material intuition, the interface includes built-in live visualizers for attention maps and neuronal activations, allowing artists to directly monitor the internal state of the network as they manipulate it.

\subsection{Levers of Control}
\label{sec:levers}

The interface exposes a suite of modulations, organized around the three components of the DiT block: self-attention, cross-attention, and the feed-forward network (Fig. \ref{fig:interface}).

\textbf{Self-Attention.} The self-attention section exposes levers that directly influence the core attention calculation. Users can apply amplitude scaling or inject noise into the attention maps (\textit{where} the model attends) or into the attention
values (\textit{what} information is aggregated). Additional controls allow for amplification or noising of the final attention outputs before they pass to the next layer.

\textbf{Cross-Attention.} Cross-attention supports the same levers as self-attention, but with two important additions. First, because cross-attention maps connect the prompt to the latent, their amplitude and noise controls can be targeted at specific prompt tokens (Section~\ref{sec:targetting}). Second, building on the methodology of the original \textit{AttentionBender}~\cite{adamcoleAttentionBenderManipulatingCrossAttention2026}, cross-attention maps can be reshaped from a flat sequence back into 3D video latents, enabling a suite of spatial-temporal modulations (rotate, scale, translate, flip, blur, and sharpen) that impose geometric and compositional constraints on the generated video, independent of the prompt.

\textbf{Feed-Forward Network (FFN).} We introduce novel controls for all stages of the sequential FFN: inputs, hidden layers, and outputs. For the hidden layers specifically, users can modulate the activation distributions via gain (amplification), thresholding (clamping low activations), and noise injection.

\subsection{Granular Network Targeting}
\label{sec:targetting}
An essential design philosophy of this tool is enabling manipulation down to the most granular components of the network. Global modifications often destroy the latent outright; targeted intervention is what makes specific layer regions and their influence on the output legible---building material intimacy with the model.

For all levers described above, users can constrain their modulations by \textbf{diffusion step} (allowing them to target coarse structural formation early in the diffusion process versus fine detail refinement later) and by \textbf{DiT layer} (allowing them to target specific depths of the transformer). Crucially, these targeting dimensions compose: any lever in Section~\ref{sec:levers} can be scoped to a specific step range, layer range, and, as described below, token or neuron band.

\begin{figure}[h!]
\centering
\includegraphics[width=1\linewidth]{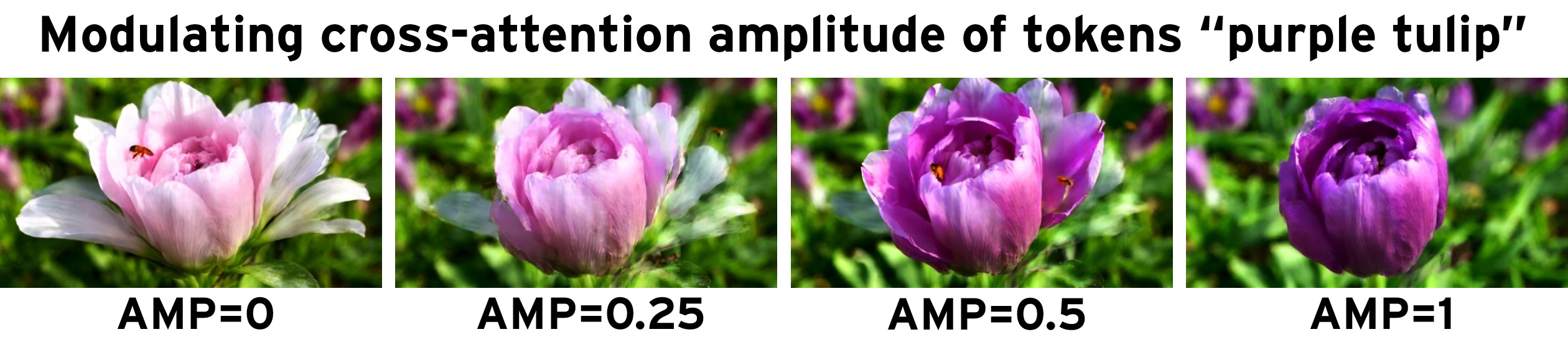}
\caption{Prompt Token Targeting: Granular control over output features is achieved by targeting specific tokens. Here, the cross-attention amplitude for the tokens ``purple tulip'' within the prompt ``a purple tulip flower'' is modulated between 0 to 1, altering the flower's color in real-time while preserving the global composition.}
\label{fig:token_target}
\end{figure}

\begin{figure*}[t]
\centering
\includegraphics[width=1\textwidth]{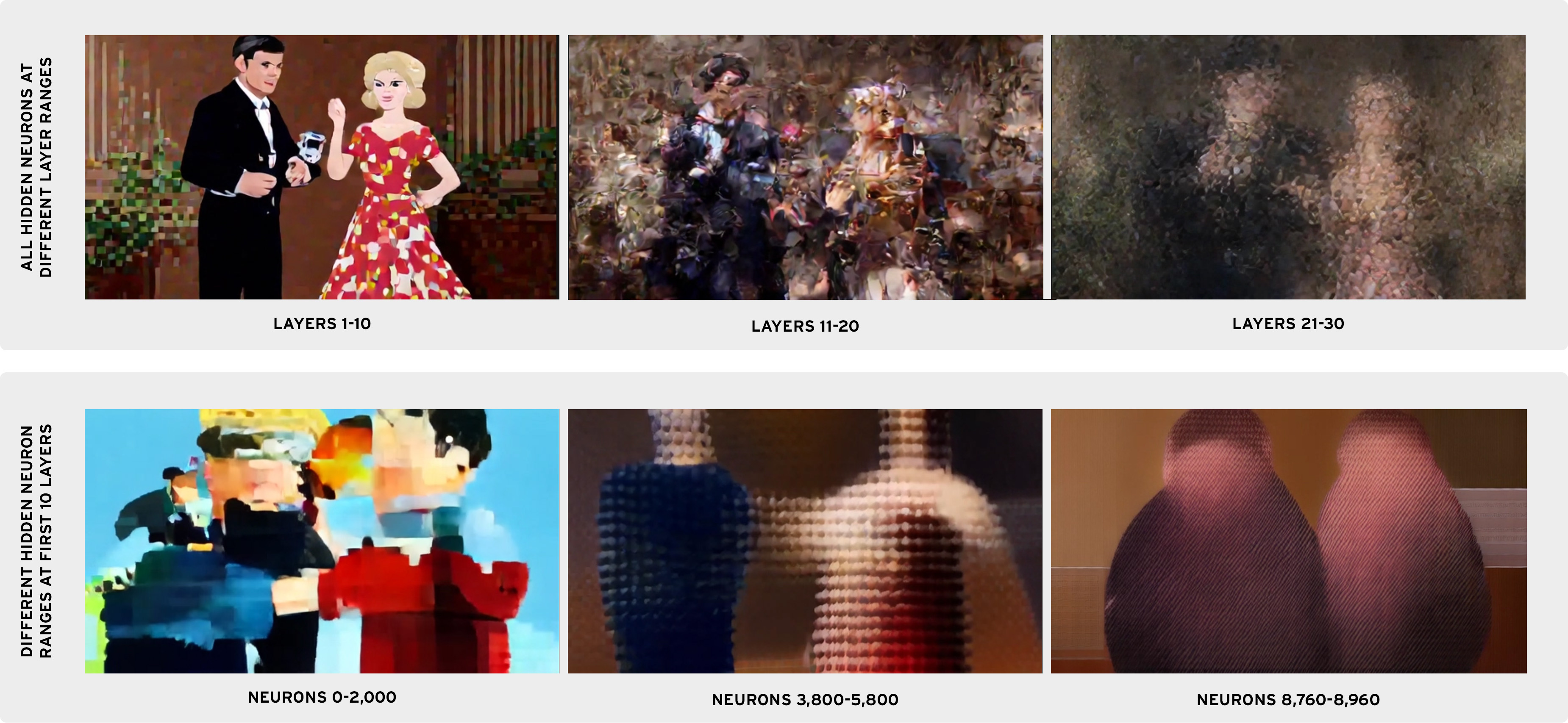}
\caption{Granular FFN Modulation: Diverse outputs generated from identical prompts using targeted hidden neuron modulation. Top row: Injecting the same amount of noise into all hidden neurons at early (0–-10), middle (10–-20), and late (20–-30) layers produces varied textural impacts, ranging from coarse geometry to fine, high-frequency details. Bottom row: Amplifying distinct ranges of hidden neurons ($3{\times}$) within the first 10 DiT layers produces varied structural interpretations. These interventions hint at the underlying functional roles of these specific network regions.}
\label{fig:neuron_target}
\end{figure*}

Furthermore, we implement domain-specific targeting. In the cross-attention layers, users can target modulations by specific \textbf{prompt tokens}. For example, an artist can isolate the tokens ``purple tulip'' in the prompt ``a purple tulip flower'' and selectively modulate its amplitude in real-time to shift the color and shape of the flower without altering the surrounding composition (Fig.~\ref{fig:token_target}). In the feed-forward layers, we enable targeting down to \textbf{individual hidden neurons} (indices 0--8,960), allowing for localized structural and textural disruptions (Fig.~\ref{fig:neuron_target}).

\section{Reflections \& Future Work}

\subsection{Expressive Potential}

\textit{Real-Time AttentionBender} has proven to be a highly expressive instrument in our early explorations. As illustrated in Figure~\ref{fig:teaser}, manipulating the DiT block yields an expansive aesthetic territory, arguably moving beyond the model's default representational tendencies. The exact same prompt and generation settings can serve as the starting point for a surprisingly diverse range of outputs. Crucially, these results feel directly rooted in the native materiality of the neural network itself, rather than mimicking the analog or digital glitch aesthetics of previous media technologies. The images presented in this paper represent only the tip of the iceberg; the full expressive range of these combinatory levers opens rich research pathways for artists and technologists alike.

\subsection{Material Intimacy as XAIxArts}

Beyond its utility as a generative tool, \textit{Real-Time AttentionBender} successfully functions as an XAIxArts probe. While prior offline network bending approaches are often limited by a cycle of parameter tweaking and waiting, bridging the gap to real-time interaction shifts the relationship we have with the tool. The ability to instantly see the impact of shifting a slider, modulating a specific layer, or targeting a token allows the artist to develop what we call a \textit{material intimacy} with the system. It transforms the abstract ``black box'' of the transformer into a visible, tactile, responsive medium. The artist may begin to ``feel'' the structure of the network, building an instinctual, embodied understanding of how a sequence of attention blocks can develop varied visual representations.

\subsection{Limitations \& Future Work}

The pursuit of real-time video diffusion introduces specific technical limitations that must be addressed in future iterations. Most notably, real-time architectures rely heavily on Key-Value (KV) caching to reduce redundant computation per step. However, because the KV cache stores the activation states of previously generated frames, it inherently resists sudden structural changes. If a bending parameter is altered mid-generation, the cached visual history can mute or override the new modulation, reducing responsiveness. Currently, we mitigate this by implementing manual ``reset cache'' and ``disable cache'' controls within the interface, which improves immediate explainability but artificially limits the length of coherent video generation. Developing a more sensitive, context-aware cache management system is a primary goal for future technical work.

Furthermore, while this paper establishes the technical architecture and initial expressive potential of the system, rigorous evaluation remains. In our immediate future work, we intend to conduct formal user studies with working media artists to evaluate the interface's expressive potential and observe how real-time network bending integrates into experimental creative workflows. Additionally, we plan to leverage this real-time environment to conduct quantitative mechanistic interpretability studies, systematically mapping the functional roles of specific DiT layers and neuronal bands to their corresponding visual outputs. As the barrier to entry for real-time models continues to drop, tools like \textit{Real-Time AttentionBender} will become increasingly vital in opening up the internal mechanics of AI to a larger audience, allowing artists to actively shape the medium rather than passively being shaped by it.


\printbibliography

\appendix

\end{document}
\endinput